# Correlation-driven topological transition in Janus VSiGeP$_2$As$_2$

Ghulam Hussain[1], Amar Fakhredine[2], Rajibul Islam[1], Raghottam M. Sattigeri[1], Carmine Autieri[1,*] and Giuseppe Cuono[1]

[1] International Research Centre MagTop, Institute of Physics, Polish Academy of Sciences, Aleja Lotników 32/46, PL-02668 Warsaw, Poland

[2] Institute of Physics, Polish Academy of Sciences, Aleja Lotników 32/46, PL-02668 Warsaw, Poland

* Correspondence: autieri@magtop.ifpan.edu.pl

**Abstract:** The appearance of intrinsic ferromagnetism in 2D materials opens the possibility of investigating the interplay between magnetism and topology. The magnetic anisotropy energy (MAE) describing the easy axis for magnetization in a particular direction is an important yardstick for nanoscale applications. Here, the first-principles approach is used to investigate the electronic band structures, the strain dependence of MAE in pristine VSi$_2$Z$_4$ (Z=P, As) and its Janus phase VSiGeP$_2$As$_2$ and the evolution of the topology as a function of the Coulomb interaction. In the Janus phase the compound presents a breaking of the mirror symmetry, which is equivalent to having an electric field, and the system can be piezoelectric. It is revealed that all three monolayers exhibit ferromagnetic ground state ordering, which is robust even under biaxial strains. A large value of coupling J is obtained, and this, together with the magnetocrystalline anisotropy, will produce a large critical temperature. We found an out-of-plane (in-plane) magnetization for VSi$_2$P$_4$ (VSi$_2$As$_4$), while in-plane magnetization for VSiGeP$_2$As$_2$. Furthermore, we observed a correlation-driven topological transition in the Janus VSiGeP$_2$As$_2$. Our analysis of these emerging pristine and Janus-phased magnetic semiconductors opens prospects for studying the interplay between magnetism and topology in two-dimensional materials.

## 1. Introduction

Since the observation of intrinsic ferromagnetism in two-dimensional layered materials (2D) such as CrGeTe$_3$ [1] and CrI$_3$ [2], the fields of magnetism and spintronics have received tremendous research attention in the 2D limit [3-14]. The atomically thin 2D magnetic materials are considered ideal systems, where the magnetic and spin-related features can effectively be controlled and modulated via proximity effects, electric field, magnetic field, strain, defects, and optical doping [15-22]. Unlike bulk materials, where magnetic ordering is possible without magnetic anisotropy, long-range magnetic ordering in layered 2D materials is not conceivable in systems deprived of magnetic anisotropy, which is necessary to balance out thermal fluctuations [23]. Due to the fact that magnetic anisotropy is primarily caused by spin-orbit coupling (SOC) effects [24], it becomes a crucial characteristic. Furthermore, spintronic devices such as magnetic tunnel junctions and spin valves show enhanced performance based on 2D magnetic structures with substantial magnetic anisotropy [25-27]. It has been demonstrated that strain engineering is an effective method of tuning the magnetic, electronic, and optical characteristics of materials [28-33].

The recently discovered new family of 2D layered materials $MA_2Z_4$, where M, A and Z represent the transition metal atoms (Mo, W, Hf, Cr, V), IV-elements (Si, Ge), and V-elements (N, As, P), respectively [34], have sparked intense interest in different studies [35-44]. These layered materials exhibit outstanding mechanical, electronic, magnetic, and optical properties [38, 44-58]. It was shown that in the Janus phases of these compounds, the breaking of the mirror symmetry brings Rashba-type spin-splitting [59-62] and that this, together with the large valley splitting, can give an important contribution to semiconductor valleytronics and spintronics. In the present work, the structural, electronic and magnetic properties of pristine $VSi_2Z_4$ (Z=P, As) and their Janus phase $VSiGeP_2As_2$ are explored. We found ferromagnetic ordering in these systems, and their magnetic anisotropy energy (MAE) reveals a strong dependency on the biaxial strain. Besides, an out-of-plane direction is found as an easy axis for the magnetization of $VSi_2P_4$, while an in-plane direction is favored in $VSi_2As_4$ and $VSiGeP_2As_2$. In the Janus phase, the compound presents breaking of the mirror symmetry, this can give piezoelectric properties, and it is equivalent to having an electric field, which can manipulate magnetism and produce skyrmions in 2D materials [63, 64]. Intriguingly, there occurs a topological phase transition from a trivial to topologically non-trivial state in $VSiGeP_2As_2$ monolayer, when the Hubbard U parameter is increased. Our investigation of these compounds opens prospects for studying their intrinsic magnetism, the interplay between magnetism and topology in two-dimensional materials and spin control in spintronics.

## 2. Computational details

First-principles relativistic approach based on density functional theory (DFT) using Vienna *Ab Initio* Simulation Package (VASP) [65, 66] is employed. The Perdew–Burke–Ernzerhof (PBE) formalism in the framework of generalized gradient approximation (GGA) was used to include the electron exchange-correlation [67]. Also, we implemented the projector-augmented wave scheme to resolve the Kohn-Sham equations through the plane-wave basis set. An energy cutoff of 500 eV is taken into account for the expansion of wave functions. The Monkhorst–Pack scheme is applied for *k*-point sampling with 15×15×1 *k*-point mesh. The lattice constants were optimized at the PBE level. The optimized lattice constant for the Janus $VSiGeP_2As_2$ structure is 3.562 Å, which is between those of $VSi_2P_4$ (3.448 Å) and $VSi_2As_4$ (3.592 Å) monolayers. In addition, the convergence criterion for force is taken as 0.0001 eV/Å, while $10^{-7}$ eV of energy tolerance is considered for the lattice relaxation. Also, the number of electrons treated as valence is 41. In examining the dynamical stability, a 4×4×1 supercell of $VSiGeP_2As_2$ monolayer is taken for calculating the phonon dispersion using the PHONOPY code [68]. The GGA+U routine, along with SOC, is executed, and the strongly correlated correction intended for V-*3d* is considered throughout the calculations. The values of the Hubbard parameter used for the d-orbitals of V are U = 4 eV for $VSi_2P_4$, and 2 eV for $VSi_2As_4$ and $VSiGeP_2As_2$, and the Hund coupling $J_H$ is set at 0.87 eV. The main source of SOC in this compound is As; the value of SOC for As was estimated to be 0.164 eV [69, 70].

3. **Results and discussion**

The monolayered $VSi_2Z_4$ (Z=P, As) 2D materials crystallize in a hexagonal geometry with $P\bar{6}m2$ (No. 187) as the space group. These structures are seven-atom thick monolayered systems; the atoms are strongly bonded together with the order as Z-Si-Z-V-Z-Si-Z for pristine and P-Si-P-V-As-Ge-As in the case of the Janus phase. Figure 1a shows the pristine $VSi_2P_4$, $VSi_2As_4$, and Janus $VSiGeP_2As_2$ structures. The $VSi_2Z_4$ (Z=P, As) monolayers have broken inversion symmetry while protecting the mirror-plane symmetry with respect to V plane. Also, the primitive cell with side and top views is shown for the Janus $VSiGeP_2As_2$ phase in Fig. 1a, which presents the breaking of mirror symmetry with regard to the V atom. This is equivalent to an electric field, and the system can show piezoelectricity. The optimized lattice constants for $VSi_2P_4$ and $VSi_2As_4$ monolayers are 3.448 Å and 3.592 Å, respectively, whereas, for the Janus $VSiGeP_2As_2$ structure, it is 3.562 Å. Fig. 1b presents the 2D Brillouin zone with the high-symmetry points indicated by red letters. Fig. 1c shows the schematic representation for the topological transition as a function of onsite Coulomb interaction in $VSiGeP_2As_2$ monolayer.

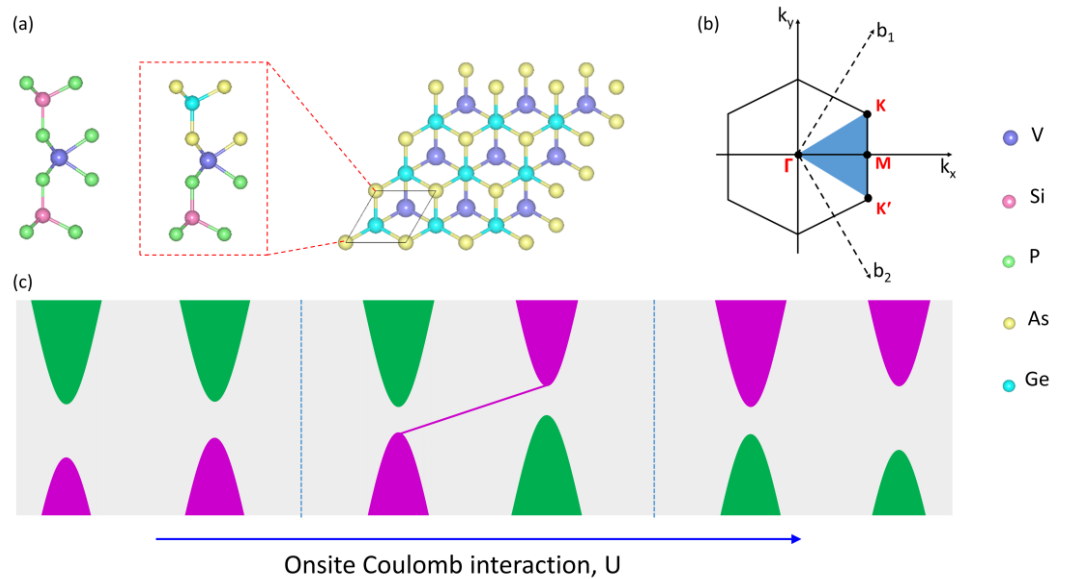

**Figure 1** (a) Side view of $VSi_2P_4$ monolayer, side and top views for Janus phase $VSiGeP_2As_2$ primitive cell. (b) 2D Brillouin zone with the high-symmetry points indicated by red letters. (c) Schematic representation for the topological phase transition as a function of onsite Coulomb interaction observed in $VSiGeP_2As_2$ monolayer.

The stabilities of pristine VSi$_2$Z$_4$ (Z=P, As) monolayers and Janus VSiGeP$_2$As$_2$ structure were studied through the cohesive energies and the phonon dispersion. We computed the cohesive energies per atom ($E_c$); for VSi$_2$Z$_4$, $E_c = [E_{VSi_2Z_4} − (E_V + 2E_{Si} + 4E_Z)]/7$, where the energy terms $E_{VSi_2Z_4}$, $E_V$, $E_{Si}$, $E_Z$ represent the total energies of VSi$_2$Z$_2$ monolayer and that of V, Si and Z atoms, respectively. Similarly, for the Janus VSiGeP$_2$As$_2$, it can be written as $E_c = [E_{VSiGeP_2As_2} − (E_V + E_{Si} + E_{Ge} + 2E_P + 2E_{As})]/7$. The values of $E_c$ were calculated as -3.25, -2.60 and -2.92 eV/atom for VSi$_2$P$_4$, VSi$_2$As$_4$ and VSiGeP$_2$As$_2$. These are relatively higher as compared to recently reported MoSiGeP$_2$As$_2$ (-2.77 eV/atom), WGeSiP$_2$As$_2$ (-2.84) [62], and other transition-metal based 2D Janus materials such as MoSSe, WSSe (−2.34 eV, -2.06 eV) [71]. Here, the phonon dispersion for VSiGeP$_2$As$_2$ is calculated along the high symmetry directions of the Brillouin zone (K-Γ-M-K) with the method of finite difference implemented in the Phonopy code. In Fig. 2a, we show the phonon dispersion of VSiGeP$_2$As$_2$ revealing no imaginary frequency modes, thus dynamically stable. The pristine monolayers VSi$_2$Z$_4$ (Z=P, As) are already reported to be dynamically stable [9, 31]. The large values of cohesive energies $E_c$, and the dynamical stability established from phononic spectra can promise their experimental realization.

The electronic configuration for an unbonded V atom is 3d$^3$4s$^2$. However, the V atom in VSi$_2$Z$_4$ (Z=P, As) is trigonal-prismatically coordinated with six Z atoms. Such type of crystal field divides the *3d* orbitals into $d_{z^2}$, $d_{yz}/d_{xz}$, and $d_{xy}/d_{x^2-y^2}$ as reported in MoS$_2$ for Mo atoms, which requires that $d_{z^2}$ orbital should be occupied first [72]. The V atom donates four electrons to neighboring Z atoms, with one electron remaining, giving rise to V$^{4+}$ valence state. With this one unpaired electron in $d_{z^2}$ a magnetic moment of 1 $\mu_B$ is expected according to Hund's rule and Pauli exclusion principle. Our DFT calculations indeed revealed a magnetic moment of ~1 $\mu_B$ per formula unit for VSi$_2$Z$_4$ (Z=P, As) and Janus VSiGeP$_2$As$_2$ structures. In addition, the total energies of two distinct magnetic configurations were evaluated in order to determine the magnetic ground state. For the antiferromagnetic (AFM) configuration, the magnetic moments were made antiparallel to nearest neighbors, instead, all magnetic moments were initialized in the same direction in the ferromagnetic (FM) configuration. In both instances, the spin orientations were off-plane. Fig. 2b depicts these two common magnetic orderings with a 2×2×1 supercell for which the total energies and magnetic moments of FM and AFM configurations were calculated, respectively. For the 2×2×1 supercell, a magnetic moment of ~4.0 $\mu_B$ is revealed for both the pristine and Janus phases in the FM state, while 0 $\mu_B$ is observed with AFM alignment. Moreover, the energy difference between the FM and AFM states (E$_{FM}$ −

$E_{AFM}$) indicated negative energies, strongly suggesting intrinsic ferromagnetism in VSi$_2$Z$_4$ (Z=P, As) monolayers and their Janus structure. The optimized lattice constants a$_o$, energy difference between FM and AFM alignments and the easy axis for the magnetization for VSi$_2$Z$_4$ (Z=P, As) and Janus phase are reported in Table 1. We also computed the average electrostatic potential profiles along the z-axis for the pristine and the Janus phase. As indicated in Fig. 2c,d the profiles are symmetric for VSi$_2$Z$_4$ (Z=P, As), however in the case of Janus VSiGeP$_2$As$_2$, the calculated average electrostatic potential is rather asymmetric with a work function difference, ΔΦ of 0.35 eV (Fig. 2e).

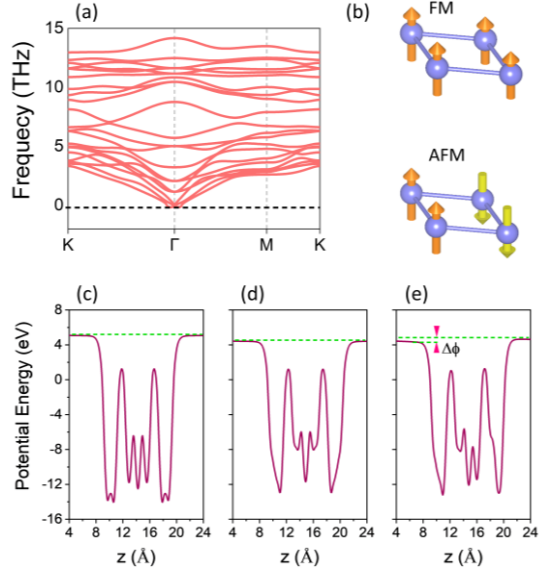

**Figure 2** (a) The phonon dispersion for Janus VSiGeP$_2$As$_2$ monolayer indicating no imaginary frequencies. (b) Two magnetic configurations FM and AFM, considered to evaluate the magnetic ground state. The planar average electrostatic potential energy of (c) VSi$_2$P$_4$, (d) VSi$_2$As$_4$, and (e) Janus VSiGeP$_2$As$_2$ monolayers. The work function difference ΔΦ is estimated to be 0.35 eV for the Janus phase.

**Table 1** Optimized lattice constants a$_o$, energy differences between FM and AFM alignments and the easy axis for the magnetization.

| Material | a$_o$ (Å) | [E$_{FM}$ − E$_{AFM}$] (eV) | Easy axis |
| --- | --- | --- | --- |
| VSi$_2$P$_4$ | 3.448 | -0.143 | Out-of-plane |
| VSi$_2$As$_4$ | 3.592 | -0.202 | Inplane |
| VSiGeP$_2$As$_2$ | 3.562 | -0.210 | Inplane |

The transition metal based 2D materials host degenerate energy valleys (at K, K' points of Brillouin zone) owing to a lack of inversion symmetry. Such energy valleys can be manipulated and utilized in valley-spin Hall effects and valley-spin locking [73-75]. Generating and controlling the valley polarization by making the K, K' valleys non-degenerate is a big challenge in valleytronics. There are multiple means to lift this valley degeneracy between the K, K' valleys and consequently generate the valley polarization. However, when an external magnetic field is removed, the polarization disappears. In general, the 2D monolayers preserve the long-range ferromagnetic ordering due to the intrinsic anisotropy. Specifically in V-based TMDs, the spontaneous valley polarization results from the magnetic interaction among the V-3d electrons, which is independent of external fields and enables the modulation of spin and valley degrees of freedom. We therefore investigate the orbital-projected band structures of $VSi_2Z_4$ (Z=P, As) and Janus $VSiGeP_2As_2$ monolayers as shown in Fig. 3. As illustrated, all the three structures reveal nondegenerate energy values at the K and K' valleys, and as a result they show different energy band gaps at the two valleys. The valley polarization is defined as [5], $\Delta E_{v/c} = E^{K'}_{v/c} - E^{K}_{v/c}$, where $E^{K,K'}_{v/c}$ represent the energies of electronic band edges at K/K' valleys, correspondingly. In the case of $VSi_2P_4$, using this definition, we found a valley polarization of 76.6 meV in the bottom conduction band, while the top valence bands at K/K' valleys remain almost degenerate with valley polarization of -3.9 meV. By contrast, for $VSi_2As_4$, the valley polarization is -8.2 meV in the bottom conduction band, whereas that of the top valence band it is calculated to be ~88 meV. On the other hand, in the Janus phase, the bottom conduction bands at K/K' remain almost degenerate in energy with valley polarization of -5 meV and 73.3 meV in the top valence bands. This reveals that intrinsic ferromagnetism is much more efficient in creating valley polarization. In addition, the conduction band minimum (CBM) in $VSi_2P_4$ is composed of V-$d_{xy}$ and V-$d_{x^2-y^2}$ states at both K and K' points, while the valence band maximum (VBM) is majorly composed of V-$d_{z^2}$ orbitals. On the other hand, this orbital composition becomes reverse for pristine $VSi_2As_4$ and Janus $VSiGeP_2As_2$ i.e. V-$d_{z^2}$ orbitals contribute to the CBM, while V-$d_{xy}$ and V-$d_{x^2-y^2}$ form the VBM.

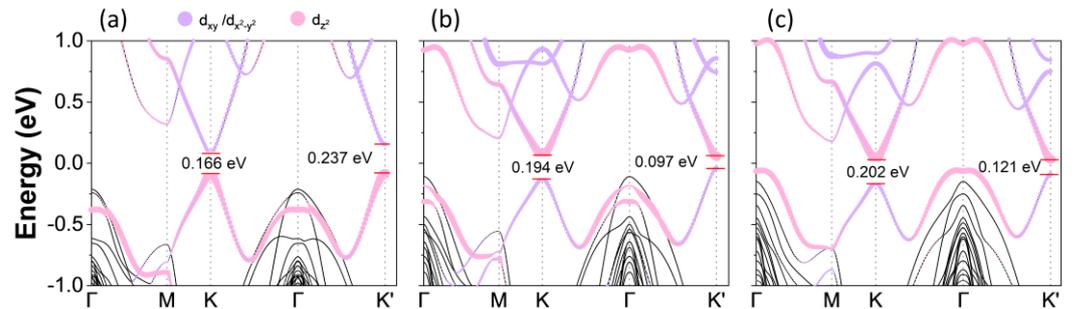

**Figure 3** Orbitally resolved electronic band structures of (a) $VSi_2P_4$, (b) $VSi_2As_4$, and (c) $VSiGeP_2As_2$ Janus structure. The V-*3d* orbitals are represented by different colors, where the size of the colored dot describes the contribution from particular orbitals. The contribution decreases as the size of the colored dot decreases. The values of the Hubbard parameter used for the d-orbitals of V are U = 4 eV for $VSi_2P_4$, and 2 eV for $VSi_2As_4$ and $VSiGeP_2As_2$.

We study the dependence of magnetic features of the VSi$_2$Z$_4$ and Janus VSiGeP$_2$As$_2$ on the biaxial strain. The energy difference between the FM and AFM configuration (E$_{FM}$ − E$_{AFM}$), which determines the magnetic ground for the material, is illustrated in Fig. 4 as a function of compressive and tensile strains. All systems retain the FM orderings under different biaxial strains and do not show any phase transition from FM to AFM state with the applied strain. The strain, in this instance, is defined as follows:

$$\varepsilon = \left(\frac{a - a_o}{a_o}\right) \times 100\%$$

Here, 'a$_o$' designates the lattice constant at a strainless state, and 'a' represents the strained lattice constant. The exchange parameter 'J', by taking into account the nearest neighbor exchange interactions, can be written as [28]:

$$J = -\left(\frac{E_{FM} - E_{AFM}}{6|\vec{S}|^2}\right)$$

Where $|\vec{S}|$ = ½, since the electronic configuration 3d$^3$4s$^2$ becomes 3d$^1$ after losing four electrons. The energy differences between the FM and AFM alignments can be easily calculated using DFT ground state formalism, which can be used to compute the Heisenberg exchange parameter 'J'. The large value of 'J', togheter with the magnetocrystalline anisotropy will produce a large critical temperature.

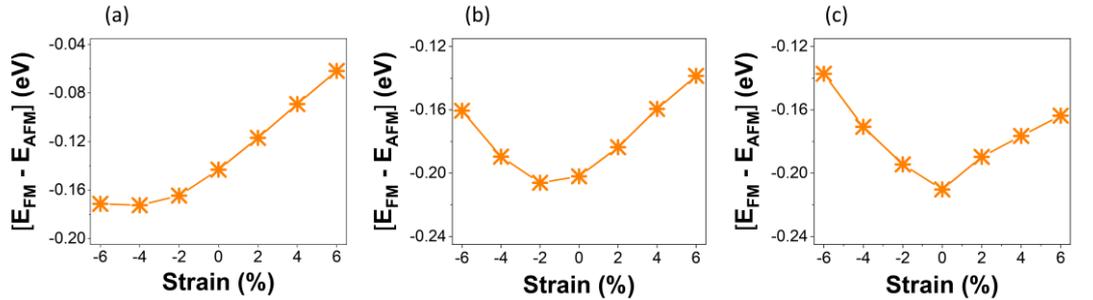

**Figure 4** Strain dependence of energy differences between two magnetic configurations (FM and AFM) for (a) VSi$_2$P$_4$, (b) VSi$_2$As$_4$, and (c) VSiGeP$_2$As$_2$ Janus structures. The values of Hubbard parameter used for the d-orbitals of V are U = 4 eV for VSi$_2$P$_4$, and 2 eV for VSi$_2$As$_4$ and VSiGeP$_2$As$_2$.

The magnetic anisotropy energy (MAE) is used to determine the easy axis for magnetization direction. It is defined as the energy difference between the out-of-plane and in-plane spin alignments i.e. MAE = E$_⊥$ - E$_∥$. Consequently, a negative MAE will indicate an out-of-plane easy-axis (perpendicular direction for magnetization), while positive values of MAE will indicate an in-plane easy-axis (magnetization parallel to the plane direction). The MAE is originated because of the reliance of magnetic attributes on a specific crystallographic direction. Classically, dipole–dipole interactions are believed to be the origin of MAE, nonetheless quantum mechanically, the main cause lies in SOC [29]. For that reason, SOC effects should be considered in the evaluation of MAE. Thus, non-collinear calculations with SOC taken into account are carried out to evaluate the total energies ($E_\perp$, $E_\parallel$) for the corresponding magnetization directions. We found MAE values of -4 μeV for $VSi_2P_4$ and 53 μeV in $VSi_2As_4$, indicating out-of-plane and in-plane magnetizations, respectively. Similarly, an in-plane magnetization is confirmed in ViSiGeP$_2$As$_2$ with an MAE value of 48 μeV. The direction of magnetization is essential to attain spontaneous valley polarization [76]. The effect of biaxial strain on MAE for all the monolayer systems is presented in Fig. 5. One can see how the MAE is influenced by the tensile and compressive strains. For $VSi_2P_4$, the MAE decreases for either strain directions, with persistent out-of-plane easy axis for magnetization as shown in Fig. 5a. On the other hand, the in-plane easy axis in $VSi_2As_4$ is found tunable, it can be transformed to out-of-plane direction by applying some critical tensile or compressive strains as indicated in Fig. 5b. Likewise, an out-of-plane magnetization can be achieved in Janus ViSiGeP$_2$As$_2$ monolayer at ε =1.5 % as shown in Fig. 5c. Shaded regions show the tuning of easy axis for the magnetization direction.

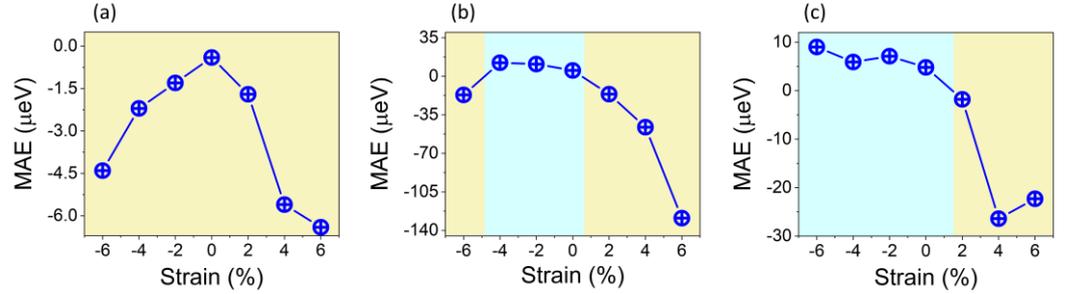

**Figure 5** MAE as a function of biaxial strain calculated for two magnetic configurations ([001], [100]) (a) VSi$_2$P$_4$, (b) VSi$_2$As$_4$, and (c) VSiGeP$_2$As$_2$ Janus structures. Shaded regions indicate the modulation of the easy axis. The brown region is for an out-of-plane easy-axis, while the cyan region indicates an in-plane easy-axis. The values of the Hubbard parameter used for the d-orbitals of V are U = 4 eV for VSi$_2$P$_4$, and 2 eV for VSi$_2$As$_4$ and VSiGeP$_2$As$_2$.

Next, we show the electronic band structures of Janus ViSiGeP$_2$As$_2$ monolayer by varying onsite Coulomb interaction known as the Hubbard parameter 'U' and by taking the SOC effect in consideration. Clearly, the CBM at the K/K' valleys is made up of V-$d_{z^2}$ orbitals when U=2 eV is in the strain-free state, whereas the VBM is composed of V-$d_{xy}$ and V-$d_{x^2-}$

$_{y^2}$ states. Upon increasing the Hubbard parameter 'U', the V-$dz^2$ orbitals come down in energy, while the $d_{xy}/d_{x^2-y^2}$ states go up in energy. When U reaches 2.8 eV, the system becomes gapless at the K' point, although gapped at the K valley. The gapless nature of the band structure at K' displays Weyl-like linear dispersion. Further raising the U, the electronic band gap becomes smaller and smaller at K valley. Conversely, at the K' valley the band gap opens again with a band inversion exchanging the orbital contributions of the valence and conduction bands as compared to the band structure at U=2 eV. Consequently, a topological phase transition occurs between U=2.8 and U=3.1 eV, leading to the emergence of the quantum anomalous Hall phase [5]. At U=3.1 eV, the band gap closes at the K point and starts to reopen at 3.2 eV, with another band inversion achieved at K valley. At U=3.2 eV, we have a band inversion at both K and K'; as a result, the Janus structure is restored to the trivial ferrovalley insulating phase. The orbitally-projected band structure at U=3.6 eV complies with all these behaviors. The evolution of band gaps and topological phases as a function of the Coulomb repulsion at both K/K' valleys is summarized in Fig. 6h. As indicated, the trend of band gaps at the two valleys are quite similar; they begin to diminish, then they go to zero, and finally they reopen by increasing U. Since the band gap is smaller at K' than at K valley (when U=2 eV), the critical Hubbard parameter U necessary for closing the band gap is not similar; it is U=2.8 eV and 3.1 eV, respectively. While usually the Coulomb repulsion kills the topological properties, in this case the Coulomb repulsion is necessary to observe the topological phase. Additionally, the range of U where the topological phase appears is between 2.8 and 3.1 eV, that is a realistic physical range for the Coulomb repulsion of 3d electrons. Moreover, the orbital characters at the K/K' points of the Brillouin zone are investigated as shown in schematic Fig. 7, which reveals the splitting of the energy levels of d orbitals in a trigonal prismatic crystal field environment. Here, only the middle layer containing V ions is displayed since the nonmagnetic top and bottom layers of these monolayers do not contribute to the spin density distribution.

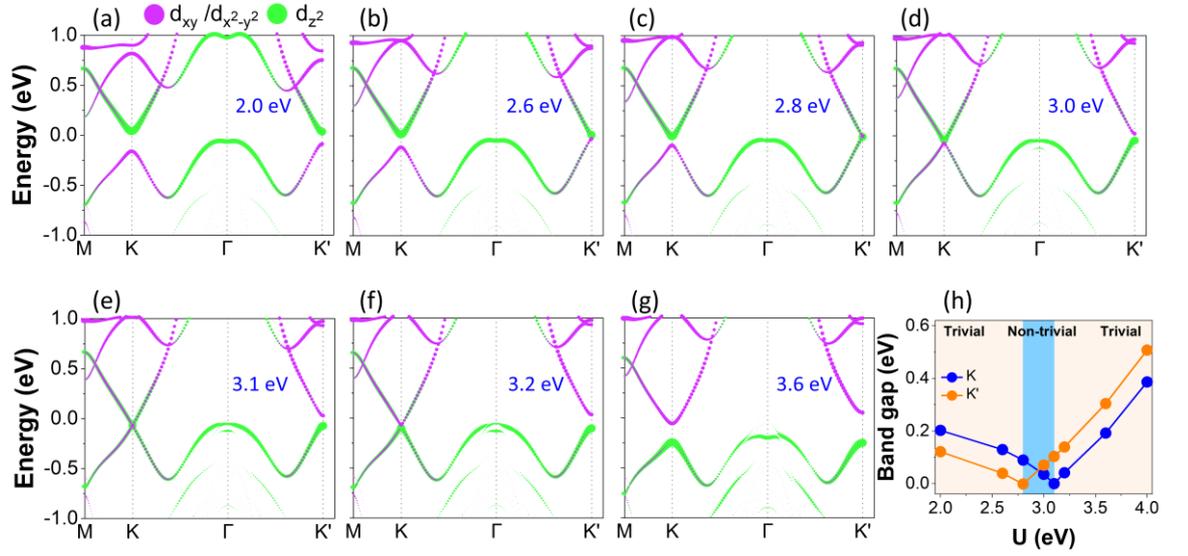

**Figure 6** (a-g) Orbitally-resolved electronic band structures of Janus VSiGeP$_2$As$_2$ monolayer with PBE+U and SOC included under different Hubbard parameter values U. The size of the colored dot is proportional to the weight of the corresponding orbitals. (h) Band gaps for the two K/K' valleys.

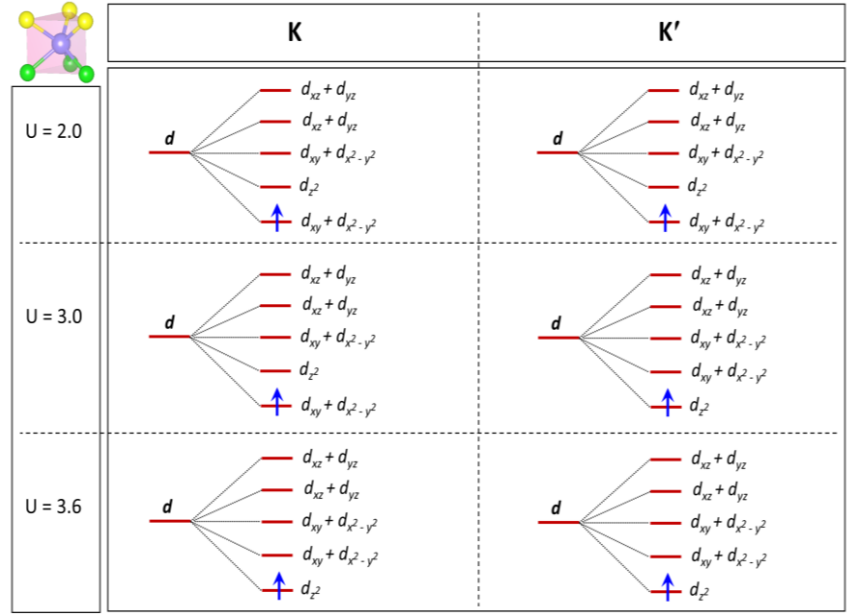

**Figure 7** A schematic for the evolution of d orbitals of the spin up-subsector as a function of Hubbard parameter U for Janus VSiGeP$_2$As$_2$ monolayer at K/K' valleys. At U=2 eV and U=3.6 eV, the system is in the trivial ferrovalley insulating phase, while at U=3 eV, it is in the topological phase.

## Conclusions

In conclusion, based on first principles calculations, we present a detailed and comprehensive study of pristine VSi$_2$Z$_4$ (Z=P, As) and Janus VSiGeP$_2$As$_2$ monolayers. In the Janus phase, the compound shows breaking of the mirror symmetry, that is equivalent to having an electric field, and the system can be piezoelectric. After exploring their structural stability through ground state energies and phononic spectra, the electronic, magnetic and topological features are investigated. It is observed that these structures exhibit ground-state ferromagnetic ordering that persists at any tensile and compressive strains. In addition, VSi$_2$P$_4$ shows -4 μeV MAE with out-of-plane easy-axis, which increases with the atomic number of pnictogens; for instance, in VSi$_2$As$_4$ the MAE increases dramatically to 53 μeV with in-plane magnetization direction. Likewise, an in-plane magnetization is established in VSiGeP$_2$As$_2$ with an MAE value of 48 μeV. Besides, we analyzed the effect of strain on the magnetic properties such as MAE, which revealed strong dependence on the biaxial strain.

We investigated how the topology of VSiGeP$_2$As$_2$ evolves as a function of the Coulomb interaction, and we observed the topological phase in the physical range of Hubbard U for 3d electrons. Our analysis of these emerging pristine and Janus-phased magnetic semiconductors opens prospects for studying the interplay between magnetism and topology in two-dimensional materials.

**Author Contributions:** Conceptualization, C.A., G. C., G. H.; Data curation, G.H.; Investigation, G.H.; Methodology,C. A., G.C., G. H.; Writing—original draft, C. A., G. C., G. H.;Writing—review & editing, C. A., G. C., G. H., A. F., R. I., R. S., supervision C. A., G. C. All authors have read and agreed to the published version of the manuscript

**Funding:** The work is supported by the Foundation for Polish Science through the International Research Agendas program co-financed by the European Union within the Smart Growth Operational Programme (Grant No. MAB/2017/1).

**Institutional Review Board Statement:** Not applicable.

**Informed Consent Statement:** Not applicable.

**Data Availability Statement:** The data that support the findings of this study are available on request from the corresponding author

**Acknowledgments:** This work is supported by the Foundation for Polish Science through the international research agendas program co-financed by the European Union within the smart growth operational program (Grant No. MAB/2017/1). A.F. was supported by the Polish National Science Centre under Project No. 2020/37/B/ST5/02299. We acknowledge the access to the computing facilities of the Interdisciplinary Center of Modeling at the University of Warsaw, Grants No. G75-10, No. GB84-0, No. GB84-1 and No. GB84-7. We acknowledge the access to the computing facilities of the Poznan Supercomputing and Networking Center Grant No. 609.

**Conflicts of Interest:** The authors declare no conflict of interest.

1. Gong, C., et al., *Discovery of intrinsic ferromagnetism in two-dimensional van der Waals crystals.* Nature, 2017. **546**(7657): p. 265-269.
2. Huang, B., et al., *Layer-dependent ferromagnetism in a van der Waals crystal down to the monolayer limit.* Nature, 2017. **546**(7657): p. 270-273.
3. Li, Y., et al., *Modulated Ferromagnetism and Electric Polarization Induced by Surface Vacancy in MX2 Monolayers.* The Journal of Physical Chemistry C, 2022.
4. Tiwari, S., et al., *Computing Curie temperature of two-dimensional ferromagnets in the presence of exchange anisotropy.* Physical Review Research, 2021. **3**(4): p. 043024.
5. Sheng, K., et al., *Strain-engineered topological phase transitions in ferrovalley 2 H− RuCl 2 monolayer.* Physical Review B, 2022. **105**(19): p. 195312.
6. Li, C. and Y. An, *Tunable magnetocrystalline anisotropy and valley polarization in an intrinsic ferromagnetic Janus 2 H-VTeSe monolayer.* Physical Review B, 2022. **106**(11): p. 115417.
7. Yin, Y., et al., *Emerging versatile two-dimensional MoSi $ _2 $ N $ _4 $ family.* arXiv preprint arXiv:2211.00827, 2022.
8. Islam, R., et al., *Tunable spin polarization and electronic structure of bottom-up synthesized MoSi 2 N 4 materials.* Physical Review B, 2021. **104**(20): p. L201112.
9. Feng, X., et al., *Valley-related multiple Hall effect in monolayer V Si 2 P 4.* Physical Review B, 2021. **104**(7): p. 075421.
10. Autieri, C., et al., *Limited Ferromagnetic Interactions in Monolayers of MPS3 (M= Mn and Ni).* The Journal of Physical Chemistry C, 2022. **126**(15): p. 6791-6802.
11. Basnet, R., et al., *Controlling magnetic exchange and anisotropy by non-magnetic ligand substitution in layered MPX3 (M= Ni, Mn; X= S, Se).* Phys. Rev. Research, 2022. **4**,( 023256).
12. Liu, W., et al., *A three-stage magnetic phase transition revealed in ultrahigh-quality van der Waals bulk magnet CrSBr.* ACS nano, 2022. **16**(10): p. 15917-15926.
13. Liu, X., et al., *Canted antiferromagnetism in the quasi-one-dimensional iron chalcogenide BaFe 2 Se 4.* Physical Review B, 2020. **102**(18): p. 180403.
14. Liu, P., et al., *Strain-driven valley states and phase transitions in Janus VSiGeN4 monolayer.* Applied Physics Letters, 2022. **121**(6): p. 063103.
15. Deng, Y., et al., *Gate-tunable room-temperature ferromagnetism in two-dimensional Fe3GeTe2.* Nature, 2018. **563**(7729): p. 94-99.
16. Gibertini, M., et al., *Magnetic 2D materials and heterostructures.* Nature nanotechnology, 2019. **14**(5): p. 408-419.
17. Huang, B., et al., *Electrical control of 2D magnetism in bilayer CrI3.* Nature nanotechnology, 2018. **13**(7): p. 544-548.
18. Burch, K.S., *Electric switching of magnetism in 2D.* Nature Nanotechnology, 2018. **13**(7): p. 532-532.
19. Tian, Y., et al., *Optically driven magnetic phase transition of monolayer RuCl3.* Nano Letters, 2019. **19**(11): p. 7673-7680.
20. Zhao, Y., et al., *Surface vacancy-induced switchable electric polarization and enhanced ferromagnetism in monolayer metal trihalides.* Nano Letters, 2018. **18**(5): p. 2943-2949.
21. Song, X., F. Yuan, and L.M. Schoop, *The properties and prospects of chemically exfoliated nanosheets for quantum materials in two dimensions.* Applied Physics Reviews, 2021. **8**(1): p. 011312.
22. Jiang, S., et al., *Exchange magnetostriction in two-dimensional antiferromagnets.* Nature Materials, 2020. **19**(12): p. 1295-1299.
23. Mermin, N.D. and H. Wagner, *Absence of ferromagnetism or antiferromagnetism in one-or two-dimensional isotropic Heisenberg models.* Physical Review Letters, 1966. **17**(22): p. 1133.
24. Lado, J.L. and J. Fernández-Rossier, *On the origin of magnetic anisotropy in two dimensional CrI3.* 2D Materials, 2017. **4**(3): p. 035002.
25. Gurney, B., et al., *Spin valve giant magnetoresistive sensor materials for hard disk drives*, in *Ultrathin Magnetic Structures IV*. 2005, Springer. p. 149-175.
26. Prinz, G.A., *Magnetoelectronics.* science, 1998. **282**(5394): p. 1660-1663.

27. Kryder, M.H., *Magnetic thin films for data storage.* Thin solid films, 1992. **216**(1): p. 174-180.

28. Webster, L. and J.-A. Yan, *Strain-tunable magnetic anisotropy in monolayer CrCl 3, CrBr 3, and CrI 3.* Physical Review B, 2018. **98**(14): p. 144411.

29. Singla, R., et al., *Curie temperature engineering in a novel 2D analog of iron ore (hematene) via strain.* Nanoscale Advances, 2020. **2**(12): p. 5890-5896.

30. Hussain, G., et al., *Strain modulated electronic and optical properties of laterally stitched MoSi2N4/XSi2N4 (X= W, Ti) 2D heterostructures.* Physica E: Low-dimensional Systems and Nanostructures, 2022. **144**: p. 115471.

31. Zhang, J., et al., *Prediction of bipolar VSi 2 As 4 and VGe 2 As 4 monolayers with high Curie temperature and strong magnetocrystalline anisotropy.* Physical Review B, 2022. **106**(23): p. 235401.

32. Hussain, G., et al., *Electronic and optical properties of InAs/InAs0. 625Sb0. 375 superlattices and their application for far-infrared detectors.* Journal of Physics D: Applied Physics, 2022. **55**(49): p. 495301.

33. Dey, D., A. Ray, and L. Yu, *Intrinsic ferromagnetism and restrictive thermodynamic stability in MA 2 N 4 and Janus VSiGeN 4 monolayers.* Physical Review Materials, 2022. **6**(6): p. L061002.

34. Hong, Y.-L., et al., *Chemical vapor deposition of layered two-dimensional MoSi2N4 materials.* Science, 2020. **369**(6504): p. 670-674.

35. Cao, L., et al., *Two-dimensional van der Waals electrical contact to monolayer MoSi2N4.* Applied Physics Letters, 2021. **118**(1): p. 013106.

36. Bafekry, A., et al., *MoSi2N4 single-layer: a novel two-dimensional material with outstanding mechanical, thermal, electronic and optical properties.* Journal of Physics D: Applied Physics, 2021. **54**(15): p. 155303.

37. Mortazavi, B., et al., *Exceptional piezoelectricity, high thermal conductivity and stiffness and promising photocatalysis in two-dimensional MoSi2N4 family confirmed by first-principles.* Nano Energy, 2021. **82**: p. 105716.

38. Jian, C.-c., et al., *Strained MoSi2N4 monolayers with excellent solar energy absorption and carrier transport properties.* The Journal of Physical Chemistry C, 2021. **125**(28): p. 15185-15193.

39. Wang, Q., et al., *Efficient Ohmic contacts and built-in atomic sublayer protection in MoSi2N4 and WSi2N4 monolayers.* npj 2D Materials and Applications, 2021. **5**(1): p. 1-9.

40. Yu, J., et al., *High intrinsic lattice thermal conductivity in monolayer MoSi2N4.* New Journal of Physics, 2021. **23**(3): p. 033005.

41. Cui, Z., et al., *Tuning the electronic properties of MoSi2N4 by molecular doping: a first principles investigation.* Physica E: Low-dimensional Systems and Nanostructures, 2021. **134**: p. 114873.

42. Bafekry, A., et al., *Tunable electronic and magnetic properties of MoSi2N4 monolayer via vacancy defects, atomic adsorption and atomic doping.* Applied Surface Science, 2021. **559**: p. 149862.

43. Yao, H., et al., *Novel Two-Dimensional Layered MoSi2Z4 (Z= P, As): New Promising Optoelectronic Materials.* Nanomaterials, 2021. **11**(3): p. 559.

44. Lv, X., et al., *Strain modulation of electronic and optical properties of monolayer MoSi2N4.* Physica E: Low-dimensional Systems and Nanostructures, 2022. **135**: p. 114964.

45. Pham, D., *Electronic properties of a two-dimensional van der Waals MoGe 2 N 4/MoSi 2 N 4 heterobilayer: Effect of the insertion of a graphene layer and interlayer coupling.* RSC Advances, 2021. **11**(46): p. 28659-28666.

46. Yuan, G., et al., *Highly Sensitive Band Alignment of Graphene/Mosi2n4 Heterojunction Via External Electric Field.* Available at SSRN 4052449.

47. Cao, L., et al., *Two-dimensional van der Waals electrical contact to monolayer MoSi.* 2021.

48. Chen, R., D. Chen, and W. Zhang, *First-principles calculations to investigate stability, electronic and optical properties of fluorinated MoSi2N4 monolayer.* Results in Physics, 2021. **30**: p. 104864.

49. Bafekry, A., et al., *Band-gap engineering, magnetic behavior and Dirac-semimetal character in the MoSi2N4 nanoribbon with armchair and zigzag edges.* Journal of Physics D: Applied Physics, 2021. **55**(3): p. 035301.

50. Ray, A., et al., *Inducing Half-Metallicity in Monolayer MoSi2N4.* ACS omega, 2021. **6**(45): p. 30371-30375.
51. Ng, J.Q., et al., *Tunable electronic properties and band alignments of MoSi2N4/GaN and MoSi2N4/ZnO van der Waals heterostructures.* Applied Physics Letters, 2022. **120**(10): p. 103101.
52. Xu, J., et al., *First-principles investigations of electronic, optical, and photocatalytic properties of Au-adsorbed MoSi2N4 monolayer.* Journal of Physics and Chemistry of Solids, 2022. **162**: p. 110494.
53. Bafekry, A., et al., *Effect of electric field and vertical strain on the electro-optical properties of the MoSi2N4 bilayer: A first-principles calculation.* Journal of Applied Physics, 2021. **129**(15): p. 155103.
54. Li, Q., et al., *Strain effects on monolayer MoSi2N4: Ideal strength and failure mechanism.* Physica E: Low-dimensional Systems and Nanostructures, 2021. **131**: p. 114753.
55. Hussain, G., et al., *Exploring the structural stability, electronic and thermal attributes of synthetic 2D materials and their heterostructures.* Applied Surface Science, 2022. **590**: p. 153131.
56. Islam, R., et al., *Fast electrically switchable large gap quantum spin Hall states in MGe$_2$Z$_4$.* arXiv preprint arXiv:2211.06443, 2022.
57. Sheoran, S., et al., *Coupled Spin-Valley, Rashba Effect, and Hidden Spin Polarization in WSi2N4 Family.* The Journal of Physical Chemistry Letters, 2023: p. 1494-1503.
58. Islam, R., et al., *Switchable large-gap quantum spin Hall state in the two-dimensional M Si 2 Z 4 class of materials.* Physical Review B, 2022. **106**(24): p. 245149.
59. Van Thiel, T., et al., *Coupling charge and topological reconstructions at polar oxide interfaces.* Physical review letters, 2021. **127**(12): p. 127202.
60. Smaili, I., et al., *Janus monolayers of magnetic transition metal dichalcogenides as an all-in-one platform for spin-orbit torque.* Physical Review B, 2021. **104**(10): p. 104415.
61. Zhao, X., et al., *Topological properties of Xene turned by perpendicular electric field and exchange field in the presence of Rashba spin-orbit coupling.* Journal of Physics: Condensed Matter, 2022.
62. Hussain, G., et al., *Emergence of Rashba splitting and spin-valley properties in Janus MoGeSiP2As2 and WGeSiP2As2 monolayers.* Journal of Magnetism and Magnetic Materials, 2022. **563**: p. 169897.
63. Dou, K., et al., *Theoretical Prediction of Antiferromagnetic Skyrmion Crystal in Janus Monolayer CrSi2N2As2.* ACS nano, 2022.
64. Laref, S., et al., *Topologically stable bimerons and skyrmions in vanadium dichalcogenide Janus monolayers.* arXiv preprint arXiv:2011.07813, 2020.
65. Kresse, G. and D. Joubert, *From ultrasoft pseudopotentials to the projector augmented-wave method.* Physical review b, 1999. **59**(3): p. 1758.
66. Kresse, G. and J. Furthmüller, *Efficient iterative schemes for ab initio total-energy calculations using a plane-wave basis set.* Physical review B, 1996. **54**(16): p. 11169.
67. Perdew, J.P., K. Burke, and M. Ernzerhof, *Generalized gradient approximation made simple.* Physical review letters, 1996. **77**(18): p. 3865.
68. Togo, A. and I. Tanaka, *First principles phonon calculations in materials science.* Scripta Materialia, 2015. **108**: p. 1-5.
69. Cuono, G., et al., *Spin–orbit coupling effects on the electronic properties of the pressure-induced superconductor CrAs.* The European Physical Journal Special Topics, 2019. **228**(3): p. 631-641.
70. Wadge, A.S., et al., *Electronic properties of TaAs2 topological semimetal investigated by transport and ARPES.* Journal of Physics: Condensed Matter, 2022. **34**(12): p. 125601.
71. Li, F., et al., *Electronic and optical properties of pristine and vertical and lateral heterostructures of Janus MoSSe and WSSe.* The journal of physical chemistry letters, 2017. **8**(23): p. 5959-5965.
72. Yan, S., et al., *Enhancement of magnetism by structural phase transition in MoS2.* Applied Physics Letters, 2015. **106**(1): p. 012408.

73. Ominato, Y., J. Fujimoto, and M. Matsuo, *Valley-dependent spin transport in monolayer transition-metal dichalcogenides.* Physical Review Letters, 2020. **124**(16): p. 166803.
74. Ahammed, R. and A. De Sarkar, *Valley spin polarization in two-dimensional h− M N (M= Nb, Ta) monolayers: Merger of valleytronics with spintronics.* Physical Review B, 2022. **105**(4): p. 045426.
75. Cui, Q., et al., *Spin-valley coupling in a two-dimensional V Si 2 N 4 monolayer.* Physical Review B, 2021. **103**(8): p. 085421.
76. Liu, P., et al., *Strain-driven valley states and phase transitions in Janus VSiGeN4 monolayer.* arXiv preprint arXiv:2206.00892, 2022.